# Scaling of Geographic Space as a Universal Rule for Map Generalization

Bin Jiang, Xintao Liu and Tao Jia

Department of Technology and Built Environment, Division of Geomatics
University of Gävle, SE-801 76 Gävle, Sweden
Email: bin.jiang@hig.se, xintao.liu@hig.se, jiatao83@163.com

**Abstract**
Map generalization is a process of producing maps at different levels of detail by retaining essential properties of the underlying geographic space. In this paper, we explore how the map generalization process can be guided by the underlying scaling of geographic space. The scaling of geographic space refers to the fact that in a geographic space small things are far more common than large ones. In the corresponding rank-size plot, this scaling property is characterized by a heavy tailed distribution such as a power law, lognormal, or exponential function. In essence, any heavy tailed distribution consists of the head of the distribution (with a low percentage of vital or large things) and the tail of the distribution (with a high percentage of trivial or small things). Importantly, the low and high percentages constitute an imbalanced contrast, e.g., 20 versus 80. We suggest that the purpose of map generalization is to retain the objects in the head and to eliminate or aggregate those in the tail. We applied this selection rule to three generalization experiments, and found that the scaling of geographic space indeed underlies map generalization. We further relate the universal rule to Töpfer's radical law (or trained cartographers' decision making in general), and illustrate several advantages of the universal rule.

**Keywords:** Heavy tailed distributions, power law, principles of selection, head/tail division rule, head/tail breaks

## 1. Introduction

Map generalization is a process of producing maps at different levels of detail by retaining essential properties of the underlying geographic space. What are the essential properties? One important property is called scaling. In the physics literature, scaling is often characterized by a heavy tailed distribution such as power law, and the scaling is referred to by other terms such as scale free, fractal, scale invariance and nonlinearity. The scaling of geographic space refers to the fact that in a geographic space small objects are far more common than large ones (Jiang 2010). For example, there are far more small cities than large ones, a phenomenon characterized by Zipf's law (Zipf 1949); far more short streets than long ones (Jiang 2007); and far more small street blocks than large ones (Jiang and Liu 2012, Lämmer, Gehlsen and Helbing 2006). The phenomenon of far more small things than large ones is a de facto heavy tailed distribution that is skewed to the right (Salingaros 2005, Clauset, Shalizi and Newman 2009). A heavy tailed distribution is also called a long tail distribution, in contrast to a thin tail distribution. The thin tail distribution is often referred to as a Gaussian distribution, in which all values of a variable center around an average value or mean, forming a bell shaped distribution with two symmetric thin tails. There are many phenomena in nature and in society that exhibit a Gaussian distribution, e.g., human height, exam scores, and car speeds. A Gaussian distribution is so common it is often called a normal distribution. However, the normal distribution is not the topic of this paper. Instead we concentrate on the heavy tailed distribution, which are said to be "more normal than normal" in characterizing and understanding many natural and societal phenomena.

A heavy tailed distribution when depicted as a rank-size plot consists of the head and the tail divided at the arithmetic mean of a variable. This is stated in the head/tail division rule, i.e., "*given a variable*

*X*, if its values *x* follow a heavy tailed distribution, then the mean (*m*) of the values can divide all the values into two parts: a high percentage in the tail, and a low percentage in the head" (Jiang and Liu 2012). This head/tail division rule can be used to characterize the scaling of geographic space. For example, there are 5,242 US cities with population greater than 8000; the average size of the cities (i.e., the arithmetic mean) is 54,353, which divides all the 5,242 cities into two classes: 744 cities with population larger than the average, accounting for 14% of all the cities in the head, and 4,498 cities with population smaller than the average, accounting for 86% of all the cities in the tail. The head/tail division rule has found many applications in designing a new classification scheme, head/tail breaks, for data with a heavy tailed distribution (Jiang 2012), and in defining natural city boundaries (Jiang and Liu 2012). For example, in delineating so called natural city boundaries, the researchers partition all street blocks into two categories: those larger than the mean and those smaller than the mean. Interestingly, the smaller ones constitute urban areas or cities, while the larger ones constitute the countryside.

This study attempts to formulate a universal rule for map generalization based on the underlying scaling of geographic space. The rule is fairly straightforward. We first rank individual geographic objects according to some measure in a decreasing order, and examine whether or not the measure of the geographic objects exhibit a heavy tailed distribution. If yes, then we select those objects in the head (or equivalently eliminate those in the tail) to be represented in a small scale map. This process goes recursively until some stop condition is met (c.f., Section 3.1). We will demonstrate that the results of the selection or elimination retain or capture the underlying scaling property of geographic space at different levels of scale. The main thesis of this paper is that the scaling of geographic space constitutes a universal rule for map generalization or mapping in general. It is the underlying scaling of geographic space that makes map generalization or mapping in general possible.

We conducted three case studies in order to verify the above idea. The first case study involves the generalization of the entire Swedish street network of 160,000 streets generated from more than 600,000 arcs or street segments. From this massive data set of streets, we continuously and recursively selected those streets in the head (greater than a mean, either an average length or an average degree of connectivity; *c.f.*, Section 3.1 for more details), and generated eight levels of detail of the street network. The second case study involves the simplification of the British coastline based on the Douglas-Peucker algorithm (Douglas and Peucker 1973). Using the algorithm, one has to determine a threshold of the farthest distance from the line segment linking two end points of a line. With the head/tail division rule, we can automatically or naturally determine the threshold (which is the arithmetic mean of the distance) to generate different levels of detail of the coastline. The third case study involves generalizing a drainage network by removing less-connected streams in the tail or retaining those well-connected ones in the head. Through the case studies, we intend to illustrate the scaling of geographic space underlies map generalization.

The remainder of this paper is structured as follows. Section 2 shows the scaling property along with some mathematical processes and appeared in large geographic spaces, and introduces in an accessible manner the three heavy tailed distributions: power law, lognormal and exponential. Section 3 reports three case studies to demonstrate that the generalization processes can be effectively achieved by retaining those objects in the head and removing those in the tail. In section 4, we further elaborate that it is the scaling property that underlies the universal rule, which is superior to Töpfer's radical law for map generalization. Finally in section 5 we draw a conclusion and point to future work.

**2. Scaling, scaling patterns and mathematics for characterizing scaling**
Before introducing heavy tailed distributions for characterizing scaling, we first look in detail at the scaling or fractal patterns of some idealized mathematical processes: the Cantor set and Koch curve (Figure 1). A Cantor set is generated by recursively cutting a unit interval into three equal segments, and removing the middle one. The generated set demonstrates the scaling pattern, i.e., there are far more small segments than large ones. In each iteration, the length of the segment is just one third of the segments in the previous iteration, i.e., $l_k = \left(\frac{1}{3}\right)^k$, where $l_k$ is the segment length at the k iteration.

On the other hand, the number of segments at each iteration is 1, 2, 4, 8, 16, and 32, i.e., $2^k$, where $k = 0, 1,...,5$, as shown in Figure 1a. Given k = 5, the mean length for the total number of segments $(1+2+4+8+16+32) = 63$ is 0.043 (c.f., Table 1 for the calculation). Thus, the number of the segments larger than the mean length 0.043 is $(1+2+4) = 7$ (11% a low percentage), while the number of segments smaller than the mean length is $(8+16+32) = 56$ (89% a high percentage) (Figure 1a). The same observation can be applied to the Koch curve, which is generated by replacing each straight segment with four segments of one third the length of a previous iteration. With the Koch curve pattern, there is obviously a high percentage of shorter segments and a low percentage of larger ones (Figure 1b). The Cantor set and Koch curve represent two different typical processes of creating fractal structure: the first recursively removes substructure, while the second recursively adds substructure. The resulting fractal dimension for the two processes is therefore respectively less than 1 and greater than 1.

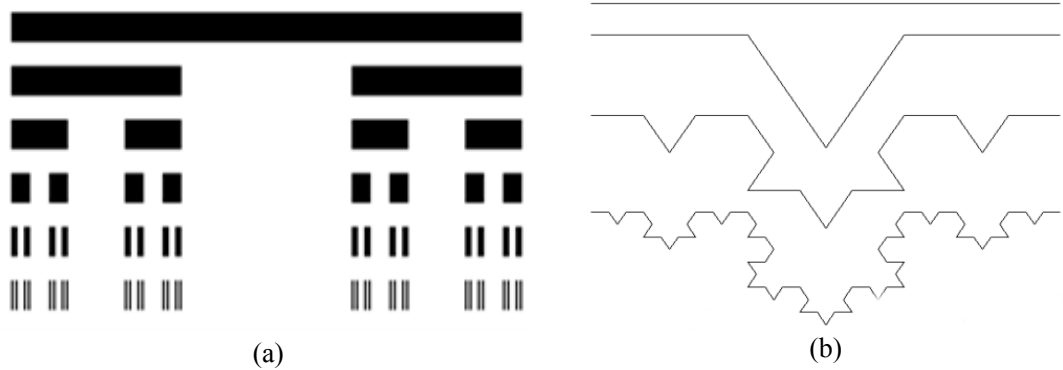

(a)　　　　　　　　　　　　　　　　　　(b)

Figure 1: Scaling patterns with (a) Cantor set, and (2) Koch curve

Table 1: Calculation of the arithmetic mean of the segment length in the Cantor set shown in Figure 1
(Note: k = iteration, l = length, # = number of segments)

| k | l | # | l*# |
|---|---|---|---|
| 0 | 1.000 | 1 | 1.000 |
| 1 | 0.333 | 2 | 0.667 |
| 2 | 0.111 | 4 | 0.444 |
| 3 | 0.037 | 8 | 0.296 |
| 4 | 0.012 | 16 | 0.198 |
| 5 | 0.004 | 32 | 0.132 |
| Sum | | 63 | 2.737 |
| Mean | | | 0.043 |

Geographic space ranging from the entire globe, to a continent, country, state, city, neighborhood, and even to a building complex essentially demonstrates the scaling property (e.g., Salingaros 2005, 2006). So do maps representing the geographic spaces; Figure 2 illustrates two such maps. The map to the left is a street map where there are many short streets (blue being the shortest) and a few long ones (red being the longest). This is the street map of the Stockholm region in Sweden. The map to the right is the geographic distribution of human settlements near Chicago (the largest patch), Illinois, USA. As we can see, there are far more small settlements than large ones. If we plot both street length and settlement sizes in a histogram or a rank-size format, we would observe a heavy tailed distribution. We leave detailed examination of heavy tailed distributions in the following case studies. It should be noted that both the histogram (or probability density function in general) and rank-size plot are consistent mathematically in showing heavy tailed distributions, but the head/tail division rule, or the head/tail breaks, adopts the rank-size plot while referring to the head and the tail.

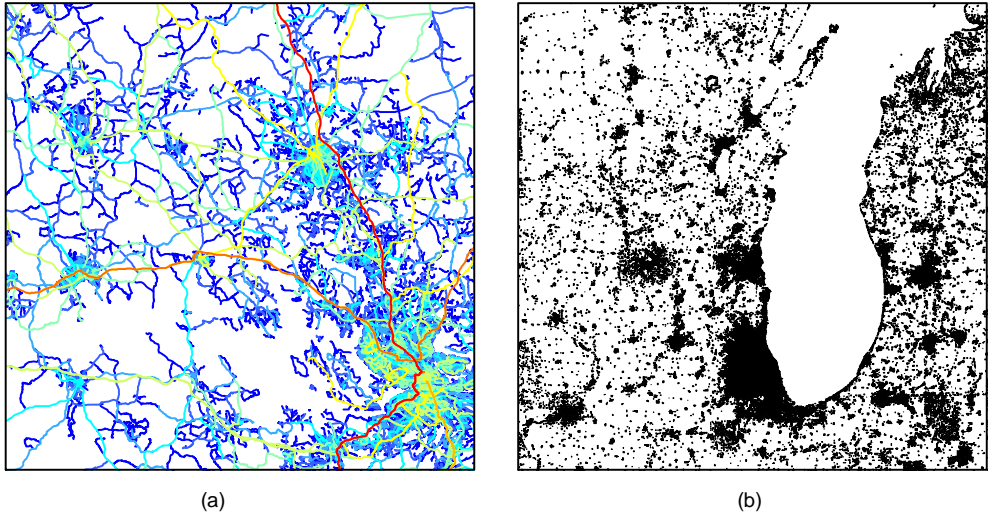

(a)  (b)

Figure 2: (Color online) Scaling of geographic space in terms of (a) street length, and (b) city sizes
(Note: A spectral color legend from blue being the shortest streets to red being the longest streets)

We now introduce three mathematical models to characterize the scaling property. It should be noted that in the literature, in particular in the literature of statistical physics, scaling refers to the phenomena of a power law distribution exclusively. However, in this paper we relax this definition to include three heavy tailed distributions, since they all have a long tail. If we let $x$ be the values of some variable (e.g., street length or city sizes), and $y$ the probability of $x$, then there are three possible relations between $x$ and $y$ that are said to be heavy tailed.

The first is the power law distribution, which is expressed by

$$y = x^{-\alpha},  \quad [1]$$

where α is called the power law exponent.

If we take the logarithm of equation [1], we get

$$\ln y = -\alpha \ln x. \quad [2]$$

It is clear that the distribution line is straight in the double logarithm plot. This straight line is often used to detect a power law through the double logarithm plot, or so called log-log plot. However, this method has been criticized for creating errors or bias in drawing a conclusion on whether a distribution is power law (Clauset, Shalizi and Newman 2009, and references therein). In the following case studies, we adopted some reliable methods based on maximum likelihood methods and the Kolmogorov-Smirnov test (Clauset, Shalizi and Newman 2009).

The second formula is the lognormal function, which has the following format:

$$y = \frac{1}{\sqrt{2\pi\sigma^2}\,x} \exp(\frac{-(\ln(x)-\mu)^2}{2\sigma^2}). \quad [3]$$

The third formula is the exponential function, which is expressed by

$$y = e^x \quad [4]$$

If we take the logarithm of equation [4], we get

$$\ln y = x \quad [5]$$

Equation [5] indicates that x and the logarithm of y have a linear relationship. Note that the constant *e* can be any other constant value, for example, in the above Cantor set, the length is decreased as an exponential function, i.e., $y = C^x$, where constant C is set to $\frac{1}{3}$.

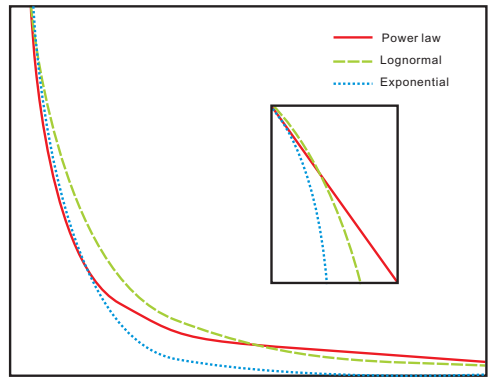

Figure 3: (Color online) Illustration of the similarity of three heavy tailed distributions in one single plot; the inset shows the same distributions in a double logarithm plot

The above three formulas with appropriate parameter settings look very similar in terms of the long tails. This similarity is illustrated in Figure 3 with an inserted log-log plot. In order to keep the mathematics at the simplest level, we have presented the above formulas in a very simple format by ignoring some constants or parameters. Interested readers may refer to the literature (e.g., Clauset, Shalizi and Newman 2009 and references therein) for more detailed descriptions. It should also be noted that with real world data it is sometimes very hard to see a power law, lognormal or exponential distribution, but their approximated versions such as a power law with exponential cutoff and stretched exponential are often appropriate (Clauset, Shalizi and Newman 2009). It should also be noted that heavy tailed distributions are sometimes defined as distributions heavier than an exponential distribution, thus excluding exponential from being a heavy tailed distribution. We believe instead that the exponential distribution can be a good approximation of heavy tailed distributions, so it is kept in the heavy tailed distribution family in this paper.

**3. A universal rule for map generalization and case studies**
In this section, we first formulate a universal rule for map generalization and then apply it to three case studies. Note that in cartography and geographic information systems (GIS) literature (e.g., Muller, Lagrange and Weibel 1995, Mackaness et al. 2007), two kinds of map generalization are defined: cartographic generalization focusing on the graphic part, and model generalization on databases. It should be noted that the universal rule can be applied to both generalization processes. The first and third case studies are for model generalization, while the second case study is for graphic generalization, i.e., to graphically represent the coast line, the point set must be reduced. The main purpose of the case studies is to illustrate the fact that the scaling of geographic space underlies map generalization or mapping in general.

**3.1 A universal rule for map generalization**
Based on the scaling of geographic space, or the head/tail breaks, we formulate a universal rule for map generalization as follows:

Rank geographic objects according to some measure in a decreasing order, and examine whether or not the measure exhibits a heavy tailed distribution. If yes, select those objects in the head to be represented in a small scale map. This process can recursively continue for the head until the objects in the head are no longer heavy tailed distributed, or are no longer a minority (e.g., > 40%).

The rule formulated above is in fact inspired by fractal or scaling patterns. For example, with the Cantor set and for the 63 segments (Figure 1), there are far more short segments than long ones. We know that fractal bears the property of self-similarity, i.e., any part is self-similar to the whole. Based on the self-similarity, it is obvious that the segments longer than the mean 0.043 are self-similar to the whole. When map scale is reduced, we take the head part to represent the whole. This process continues recursively until the head part is no longer heavy tailed distributed, or no longer a minority. This is the true meaning of self-similarity, i.e., the head part is self-similar to the whole, thus the head captures underlying pattern or structure of the whole. This generalization process is much like gradually removing the leaves of a tree (i.e., objects in the tail below the first mean), cutting out short branches step by step (i.e., objects in the tail below the second or other means), before the tree is left with a trunk only. The universality of the rule lies in the universality of the scaling of geographic space, as elaborated at the beginning of the paper. The scaling patterns are valid across every scale and every location of geographic space, and for both natural and artificial objects.

**3.2 Generalization of the Swedish street network**
We conducted the case study for the generalization of the Swedish street network using OpenStreetMap data (Bennett 2010). The street network involves 613,886 arcs or street segments, from which we generated 400,972 named streets (Jiang and Claramunt 2004) by merging all street segments with the same names. We further combined these named streets or segments without names based on good continuity according to the smallest deflection angle, and generated 166,479 natural streets (Jiang, Zhao and Yin 2008) or strokes (Thomson and Brooks 2000). The natural streets are perceptually grouped units generated from individual street segments. We simply think of the Swedish street network consisting of individual streets rather than street segments, essentially a topological way of thinking. We adopted two measures for examining scaling property: street length and street connectivity, which is the number of other streets intersected one particular street. Interestingly, the street connectivity exhibits a power law distribution, i.e., $y = x^{-3.5}$ (note: the exponent 3.5 is a bit large, usually it is between 1 and 3), and street length exhibits a lognormal distribution (Figure 4). These distributions indicate that there are far more less-connected streets than well-connected ones, or far more short streets than long ones.

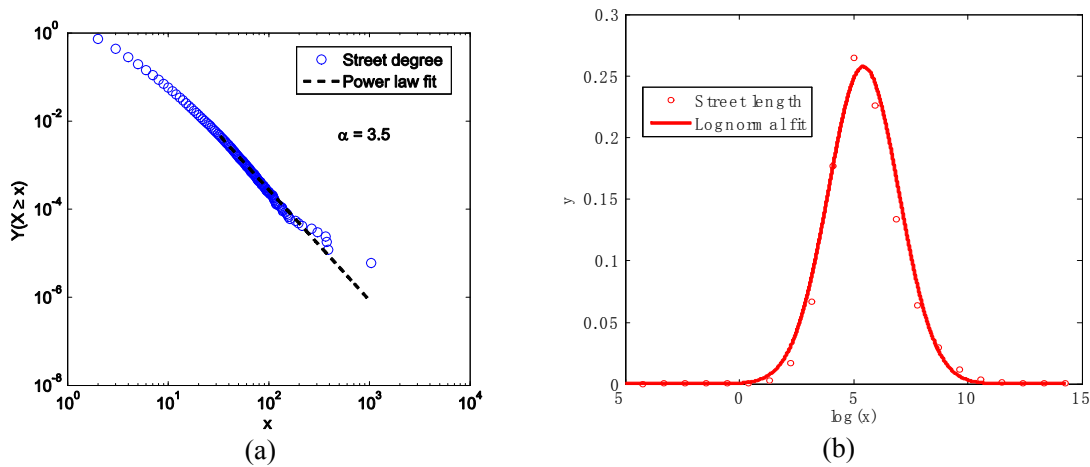

Figure 4: (Color online) Heavy tailed distributions for the street network: (a) power law distribution of the streets degree or connectivity, and (b) lognormal distribution of the streets length

Table 2: Seven levels of detail of the street network according to the streets' lengths
(Note: LOD = level of detail, # = number, % = percentage)

| LOD | # of streets | # in Head | % in Head | # in Tail | % in Tail | Mean (length) |
|---|---|---|---|---|---|---|
| Source map | 166479 | 25350 | 15% | 141129 | 85% | 1052.7 |
| Level 1 | 25350 | 5276 | 21% | 20074 | 79% | 5505.8 |
| Level 2 | 5276 | 1249 | 24% | 4027 | 76% | 17929.7 |
| Level 3 | 1249 | 309 | 25% | 940 | 75% | 45297.1 |
| Level 4 | 309 | 81 | 26% | 228 | 74% | 101694.6 |
| Level 5 | 81 | 21 | 26% | 60 | 74% | 204879.4 |
| Level 6 | 21 | 6 | 29% | 15 | 71% | 403132.5 |
| Level 7 | 6 | 2 | 33% | 4 | 67% | 747172.5 |

Given the heavy tailed distributions, and according to the suggested rule, we can use the arithmetic means to recursively select those streets in the head for the purpose of generalization. Table 2 presents the statistics on the source map and the seven levels of detail according to street lengths. Note that the head always bears a low percentage, while the tail holds a high percentage. This property remains for the source map and the derived levels of detail. The generalization can also be more effectively done according to the degree of street connectivity. Table 3 presents similar statistics with a correspondence to the generalization results shown in Figure 5. Again, there is a majority of less-connected streets in the tail and a minority of well-connected streets in the head at each level of detail. This is a clear indication of scaling property.

Table 3: Eight levels of detail (corresponding to Figure 5) of the street network according to the streets' degree or connectivity
(Note: LOD = level of detail, # = number, % = percentage)

| LOD | # of streets | # in Head | % in Head | # in Tail | % in Tail | Mean(degree) | Figure 5 |
|---|---|---|---|---|---|---|---|
| Source map | 166479 | 47021 | 28% | 119458 | 72% | 4 | Panel a |
| Level 1 | 47021 | 14382 | 31% | 32639 | 69% | 8 | Panel b |
| Level 2 | 14382 | 4235 | 29% | 10147 | 71% | 15 | Panel c |
| Level 3 | 4235 | 1294 | 31% | 2941 | 69% | 26 | Panel d |
| Level 4 | 1294 | 363 | 28% | 931 | 72% | 43 | Panel e |
| Level 5 | 363 | 101 | 28% | 262 | 72% | 72 | Panel f |
| Level 6 | 101 | 21 | 21% | 80 | 79% | 120 | Panel g |
| Level 7 | 21 | 6 | 29% | 15 | 71% | 240 | Panel h |
| Level 8 | 6 | 1 | 17% | 5 | 83% | 457 | |

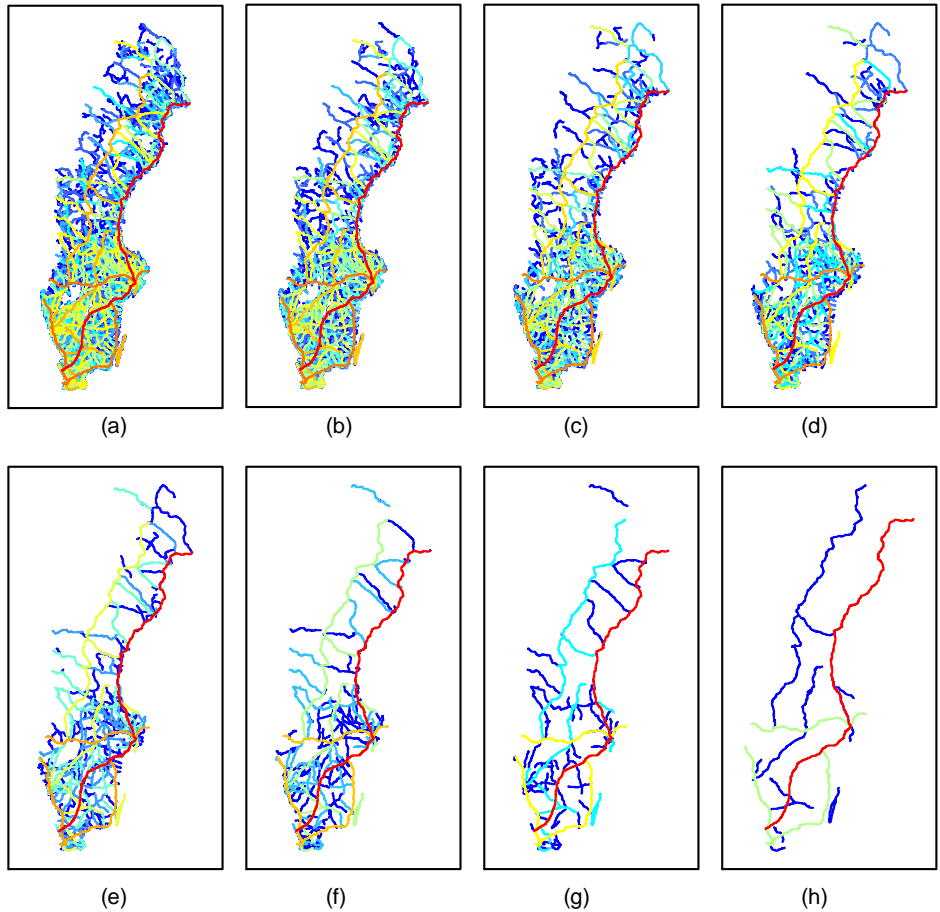

Figure 5: (Color online) The Swedish street network at different levels of detail: (a) source map, (b) level 1, (c) level 2, (d) level 3, (e) level 4, (f) level 5, (g) level 6, and (h) level 7
(Note: A spectral color legend from blue being the least connected streets to red being the most connected streets)

### 3.3 Simplification of British coastline
The second case study simplified the coastline of Britain using the Douglas-Peucker algorithm (Douglas and Peucker 1973). This algorithm is well known and can be found in many cartography or GIS textbooks. The algorithm recursively divides a curve and checks how far (x) individual points are from the line segment linking two ends of the curve. If the distance (x) is greater than a preset threshold, the points will be kept; otherwise they will be eliminated; refer to Figure 6 for an illustration. More often than not, the threshold is rather arbitrarily determined, or according to experience. We discovered that both the distance (e.g., $x_1, x_2, x_3, x_4$ and $x_5$) and ratio of distance to the length of the line segment ($x_i/d_i$) follow heavy tailed distributions. With this, we can automatically or naturally (in terms of retaining the underling scaling property) set the mean distance or the mean of the ratio as a cutoff threshold for simplification.

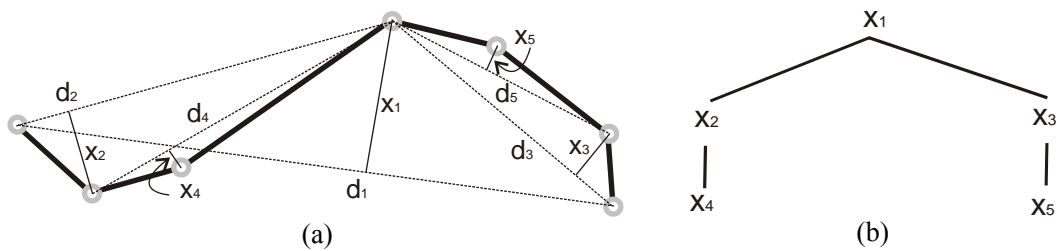

Figure 6: Illustration of Douglas-Peucker algorithm line simplification

Let us further examine how the simplification might be done. The initial British coast line contains 2,217 points. We partition this closed polygon into two pieces: a left (or west) part and a right (or east) part, each of which goes through the simplification processes. We found that both parameters $x_i$ and $x_i/d_i$ follow a lognormal distribution. As an example, Figure 7 illustrates the lognormal distribution of parameter x with the coastline curves.

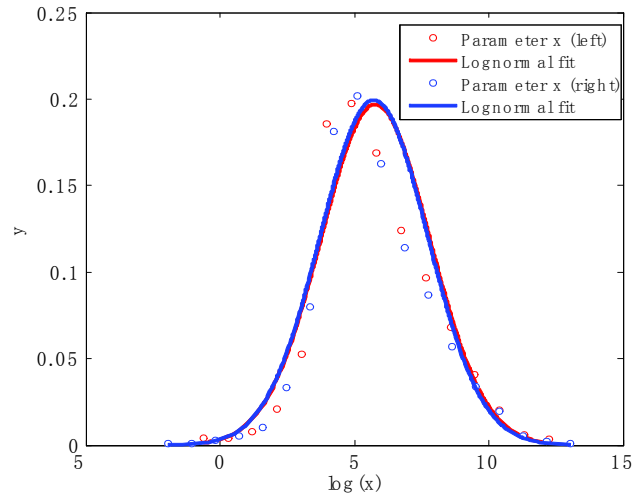

Figure 7: (Color online) Lognormal distribution of parameter x with the coastline curves

Using the universal rule, we can partition the points into two sets: a minority in the head and a majority in the tail. For simplification purposes, we simply keep those in the head while removing those in the tail (Tables 4 and 5, Figure 8). It should be noted that the percentages in the head for levels 3 and 4 are too high (>40%) to meet the condition of the heavy tailed distributions. Levels 3 and 4 are in fact invalid, for the head parts are no longer a minority. This is probably due to the small size of datasets, which we discuss further later in this paper.

Table 4: Five levels of detail of the west coastline according to the parameter (x)
(Note: LOD = level of detail, # = number, % = percentage)

| LOD | # of left points | # in Head | % in Head | # in Tail | % in Tail | Mean(x) |
|---|---|---|---|---|---|---|
| Source | 1252 | 179 | 14% | 1073 | 86% | 3062.2 |
| Level 1 | 179 | 38 | 21% | 141 | 79% | 18769.5 |
| Level 2 | 38 | 13 | 34% | 25 | 66% | 59662.4 |
| Level3 | 13 | 6 | 46% | 7 | 54% | 122689.6 |
| Level 4 | 6 | 4 | 67% | 2 | 33% | 193528.8 |

Table 5: Five levels of detail of the east coastline according to the parameter (x)
(Note: LOD = level of detail, # = number, % = percentage)

| LOD | # of right points | # in Head | % in Head | # in Tail | % in Tail | Mean(x) |
|---|---|---|---|---|---|---|
| Source | 967 | 135 | 14% | 832 | 86% | 3084.3 |
| Level 1 | 135 | 29 | 21% | 106 | 79% | 19622.3 |
| Level 2 | 29 | 9 | 31% | 20 | 69% | 65734.1 |
| Level3 | 9 | 3 | 33% | 6 | 67% | 145584.0 |
| Level 4 | 3 | 2 | 67% | 1 | 33% | 465897.4 |

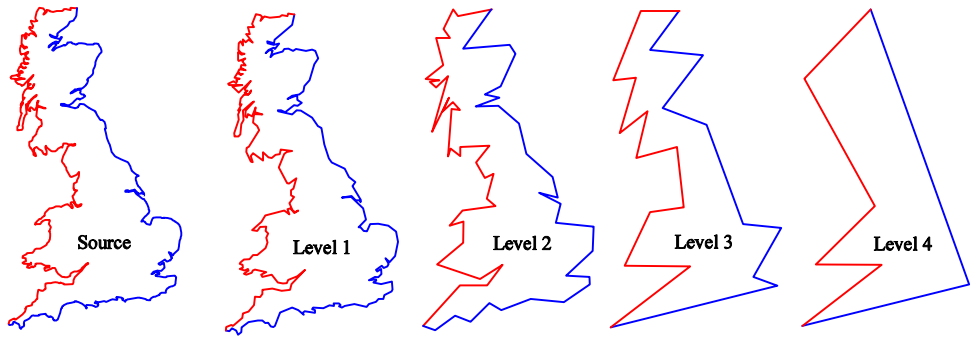

Figure 8: The British coastline at different levels of detail

We note that the levels of detail or the hierarchical levels naturally exist in data with a heavy tailed distribution. Unlike conventional generalization, where we artificially set the thresholds for the point reduction process, the thresholds here are automatically or naturally derived based on the head/tail breaks, or the underling scaling of geographic space. In fact, we identified the inherent hierarchy or levels of detail that naturally exist in the data, which can help to better guide the generalization. This is probably one of the major contributions of this paper. In this regard, this paper is not so much about a generalization technique per se, but the underlying mechanism of generalization process.

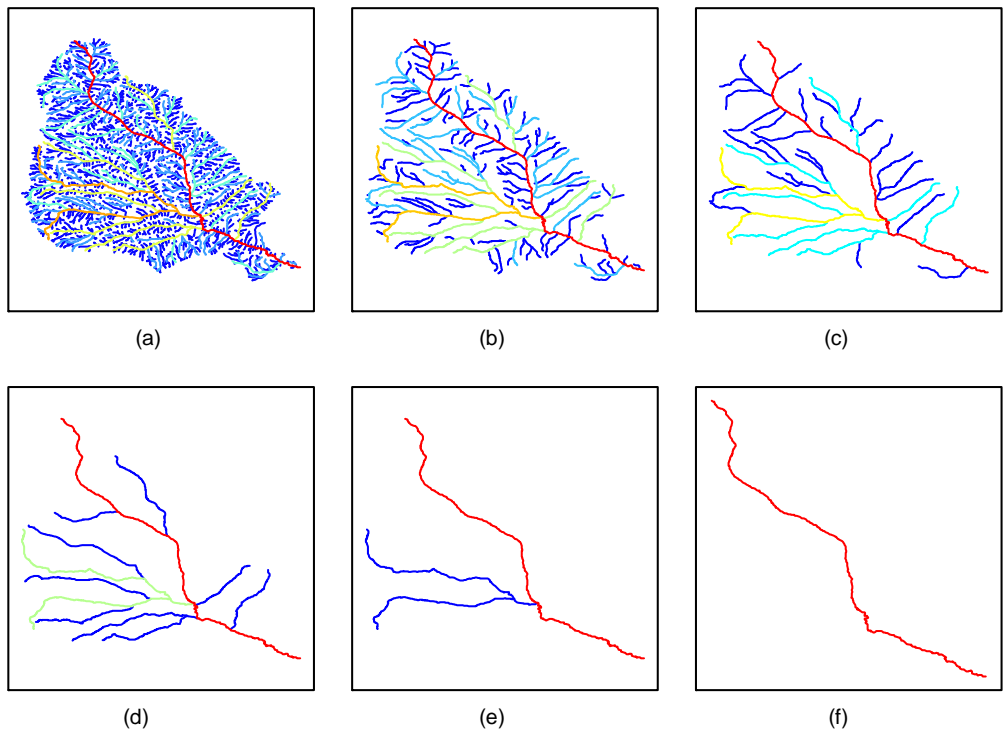

Figure 9: (Color online) The source map and five derived levels of detail of the river network
(Note: A spectral color legend from blue being the least connected streams to red being the most connected streams)

**3.4 Simplification of a drainage network**
The third case study concerns the simplification of the drainage network of the Loess Plateau in the Shanxi province of China. This network is digitized from Newman (2010), but the original data is from Pelletier (1999). The data involve 1,474 streams as shown in Figure 9a, which is visualized according to the degree of the streams' connectivity, i.e., how many other streams are connected to a stream. The visual pattern shows a similar hierarchy as the Swedish street network – far more less-connected streams than well-connected ones. The streams are formed according to some principles of perceptual grouping, the same way forming natural streets (Thomson and Brooks 2000, Jiang, Zhao and Yin 2008). The red streams are "backbones" or vital ones, while blue ones are short or trivial ones. Again, we found that the connectivity of the streams follows a power law distribution. Using the same principle of selection, we can derive different levels of detail as shown in Figure 9.

Having shown the results of the case studies, the reader may be curious about why the arithmetic mean can have such a magic effect. Putting it more specific, why should the objects larger than the mean be represented in a small scale map? According to the notion of self-similarity, any part of a scaling pattern is similar to the whole, so is the head part. There is no doubt that we should keep larger objects in a small scale map due to space limit. The question can also be addressed from the perspective of scale free, another term to refer to the scaling or fractal property. Scale free implies no scale, herein scale being size or an average size. In other words, the arithmetic mean or the average makes little sense for characterizing the sizes of power law distributed phenomena. This is in contrast to a normal distribution, where the mean can be used to characterize the values of a variable. If the values of a variable are power law distributed, the mean is not representative or a typical size of the variable. For example, we can say 1.75 meter is a typical height for human beings, but we cannot say 54,353 is a typical size of US cities (with respect to an example mentioned previously). This is because the largest city like New York can be more 300 times of the average size. However, the average value or the arithmetic mean makes a good sense in partitioning between the head and the tail. This is in line of binary thinking of the human mind, e.g., rich versus poor, large versus small, long versus short etc.. In fact, recursively applying the head/tail division rule, or the head/tail breaks, to some measure with a heavy tailed distribution leads to inherent hierarchy of the geographic objects. This is exactly the fundamental thing about map generalization.

**4. Discussions on the universal rule versus Töpfer's radical law**
Töpfer's principles of selection or the radical law (Töpfer and Pillewizer 1966), formulated based on empirical observations, is widely used to guide generalization processes (Muller, Lagrange and Weibel 1995, Mackaness et al. 2007). The radical law offers a set of mathematical expressions on how many objects from a source map should be selected to be represented in a derived small-scale map. The basic formula of the radical law is expressed as follows:

$$N_d = N_s \sqrt{M_s / M_d} \qquad [6]$$

where $N_d$ is the number of objects that are retained or selected in the derived map,

$N_s$ is the number of objects on the source map,

$M_s$ is the scale denominator of the source map, and

$M_d$ is the scale denominator of the derived map

We believe that underlying the radical law is the scaling or fractal property of geographic space. In the introduction of the paper (Töpfer and Pillewizer 1966), D. H. Maling reviewed the work by Albrecht Penck who discovered that the lengths of the coastline of Istria on maps of different scales demonstrated a heavy tailed distribution shown in Figure 1 in that paper (p. 10, Töpfer and Pillewizer 1966). This is a de fact fractal property of the coastline. Interestingly, Mandelbrot (1967) made a similar speculation on the work by Richardson (1961) to show how the lengths of geographic curves decrease as the measuring ruler increases. We see here how a cartographic insight inspired the development of fractal geometry. The radical law reflects well the insight of trained cartographers about map generalization processes. In this paper, this insight is termed as the scaling of geographic space, which should be retained at the different levels of detail of maps.

The radical law however suffers from some limitations. First, this law indicates how many objects should be selected, but it does not come up with which ones. This is a well-recognized limitation.

Second, and more critically, applying this law to generalization may end up with a biased or distorted image, since it is usually constrained by a map sheet that is subjectively or arbitrarily determined to some extent. The geographic space defined by a map sheet may not be considered to be complete. We often see that a human settlement or a hydrological network is split into map sheets. That is why with the case studies we deliberately chose some relative complete spaces: the street network for the entire country, the complete coast line, and the complete drainage network (or rive catchment). For an incomplete space, self-similarity may not hold well. People may miss the forest (i.e., the scaling pattern) for the trees (i.e., the individual things or objects) by concentrating on individual map sheets.

These two limitations of the radical law inversely show two advantages of the universal rule. First with the universal rule, it is possible to show which objects are ranked highly, and are therefore to be retained in a small scale map. Second the process of ranking and selection is based on a complete geographic space rather than the individual map sheets. On the one hand, the number of geographic objects is large enough, e.g., more than one thousand. In theory, heavy tailed distributions are for an infinite large system. The "large" is the sense of both spatial and temporal, implying that the system has a large spatial extent, and the system has sufficiently evolved temporally. On the other hand, the perspective should be in a right one, e.g., the lengths of street segments or arcs are not heavy tailed distributed, but named streets are. This indicates that it is important to have a right perspective in observing the scaling of geographic space.

Given the above argument, a good generalization strategy is to operate locally while holding an image of the scaling pattern globally. Unfortunately, many generalization studies in the literature focused on small samples with too detailed aspects, thus lacking a global picture of the underlying scaling pattern. This global scaling picture should not be constrained by the areas defined by map sheets, but rather a geographic space larger than individual map sheets. The map scales become less important in a GIS environment, since graphic manipulations such as zoom in/out make it easy to change the levels of detail. In this regard, a single detailed digital map for representing geographic space is preferred, and different levels of detail can be automatically derived (Beard 1987). We believe that the universal rule provides an effective means, more importantly a new way of thinking, to do map generalization as demonstrated in the above case studies.

The universal rule relies on the ranking or rank-size plot for the selection or elimination of geographic objects. The current ranking of geographic objects adopted in the above case studies is based on geometric or topological property only. These are just for illustration purposes. In fact, the importance or relevance of objects can be measured semantically. For example, a very small city can be more important than a large one if it bears some semantic meaning, e.g., a home town of an important person. If we could achieve the ranking effectively involving all these aspects including geometric, topological and semantic, the generalization would be more meaningful.

**5. Conclusion**
In this paper, we formulated the universal rule based on the scaling of geographic space and the head/tail breaks for map generalization. This universal rule captures the essential scaling or fractal patterns of geographic space across the different levels of detail. The different levels of detail are in fact the inherent hierarchical levels that exist with the geographic objects. In this regard, we provide an objective criterion against which generalization results can be objectively assessed or evaluated. We characterized this scaling property by three mathematical models: power law, lognormal and exponential. This is different from the existing literature, in particular in the literature of statistical physics, where scaling refers to a power law distribution only. The reason we relaxed the definition of scaling to include the other two alternatives is due to the fact that it is sometimes very difficult to detect a true power law with empirical data. Therefore we consider from a practical point of view that the two alternatives are a good approximation of the power law distribution. We have seen that the three mathematical models can be adequately justified for guiding map generalization processes because of the identification of inherent hierarchy or the different levels of detail. For geographic objects that hold a heavy tailed distribution, the generalization processes can be done simply by retaining those in the head and eliminating or aggregating those in the tail.

We further elaborated on the linkage of the universal rule and Töpfer's principles of selection, both of which are based on the scaling or fractal property of geographic space. We argued that the selection or generalization constrained by map sheets may "miss the forest for the trees". We therefore concluded that the universal rule is superior to the radical law, in particular in database generalization and multi-scale spatial representations. The case studies involve only selection or elimination for the generalization processes, but other generalization operations can be guided by the universal rule as well. This of course warrants a further study in the future.

**Acknowledgement**

We thank the editor Mei-Po Kwan, the anonymous referees, and Barbara P. Buttenfield for insightful comments that help dramatically improve the quality of this paper. We also would like to thank Petra Norlund for polishing up our English.